\documentclass[twocolumn]{my_aastex631}

\published{October 2023 • © 2023. The Author(s). Published by the American Astronomical Society.}
\submitjournal{Research Notes of the AAS}

\shorttitle{Spectral Background-Subtracted Activity Maps}
\shortauthors{Denker et al.}

\usepackage{charter}
\usepackage{amsmath}

\usepackage{xspace}
\def\Hbeta{\mbox{H$\beta$}\xspace}
\def\CaK{\mbox{Ca\,\textsc{ii}\,K}\xspace}

\begin{document}

%===============================================================================
%   Opening
%===============================================================================

\title{\Large Spectral Background-Subtracted Activity Maps}

\correspondingauthor{Carsten Denker}
\email{cdenker@aip.de}

\author[0000-0002-7729-6415]{Carsten Denker}
\affiliation{Leibniz-Institut f{\"u}r Astrophysik Potsdam (AIP), 
    An der Sternwarte 16, 
    14482 Potsdam, Germany}

\author[0000-0003-1054-766X]{Meetu Verma}
\affiliation{Leibniz-Institut f{\"u}r Astrophysik Potsdam (AIP), 
    An der Sternwarte 16, 
    14482 Potsdam, Germany}

\author[0000-0002-0484-7634]{Alexander G.M.\ Pietrow}
\affiliation{Leibniz-Institut f{\"u}r Astrophysik Potsdam (AIP), 
    An der Sternwarte 16, 
    14482 Potsdam, Germany}

\author[0000-0002-3694-4527]{Ioannis Kontogiannis}
\affiliation{Leibniz-Institut f{\"u}r Astrophysik Potsdam (AIP), 
    An der Sternwarte 16, 
    14482 Potsdam, Germany}
    
\author[0000-0003-2059-585X]{Robert Kamlah}
\affiliation{Leibniz-Institut f{\"u}r Astrophysik Potsdam (AIP), 
    An der Sternwarte 16, 
    14482 Potsdam, Germany}
\affiliation{Universit{\"a}t Potsdam,
    Institut f{\"u}r Physik und Astronomie, 
    Karl-Liebknecht-Stra{\ss}e 24/25,
    14476 Potsdam, Germany}

%===============================================================================
%   Abstract
%===============================================================================

\begin{abstract}
High-resolution solar spectroscopy provides a wealth of information from photospheric and chromospheric spectral lines. However, the volume of data easily exceeds hundreds of millions of spectra on a single observation day. Therefore, methods are needed to identify spectral signatures of interest in multidimensional datasets. Background-subtracted activity maps (BaSAMs) have previously been used to locate features of solar activity in time series of images and filtergrams. This research note shows how this method can be extended and adapted to spectral data.
\end{abstract}


%===============================================================================
%   Introduction
%===============================================================================

\section{Introduction}

Initial ideas for extracting and visualizing temporal variations in time series were presented by \citet{Verma2012}. The background-subtracted magnetic flux variation revealed a system of radial spokes in which moving magnetic features preferentially migrate, connecting a small sunspot to the adjacent supergranular cell boundary. The concept of a \textit{Background-Subtracted Activity Map} (BaSAM) has been systematically introduced by \citet{Denker2019} to study changes in the solar cycle, as seen in time series of full-disk UV images and photospheric magnetograms. Flare transients can be captured by BaSAMs \citep{Pietrow2023}, which, in combination with the Color Collapsed Plotting \citep[COCOPLOT,][]{Druett2022} technique, reveal the temporal evolution of spectral features. \citet{Kamlah2023} extended the scope of BaSAMs to examine high-spatial-resolution data of pores and light-bridges and to imaging spectroscopy of the chromospheric H$\alpha$ line. These results motivated this research note, which promotes BaSAMs as a tool for the analysis of multidimensional spectroscopic data (space, time, wavelength, and polarization state) from both ground-based solar observatories and space missions.

%===============================================================================
%   Observations
%===============================================================================

\begin{figure*}
\includegraphics[width=\textwidth]{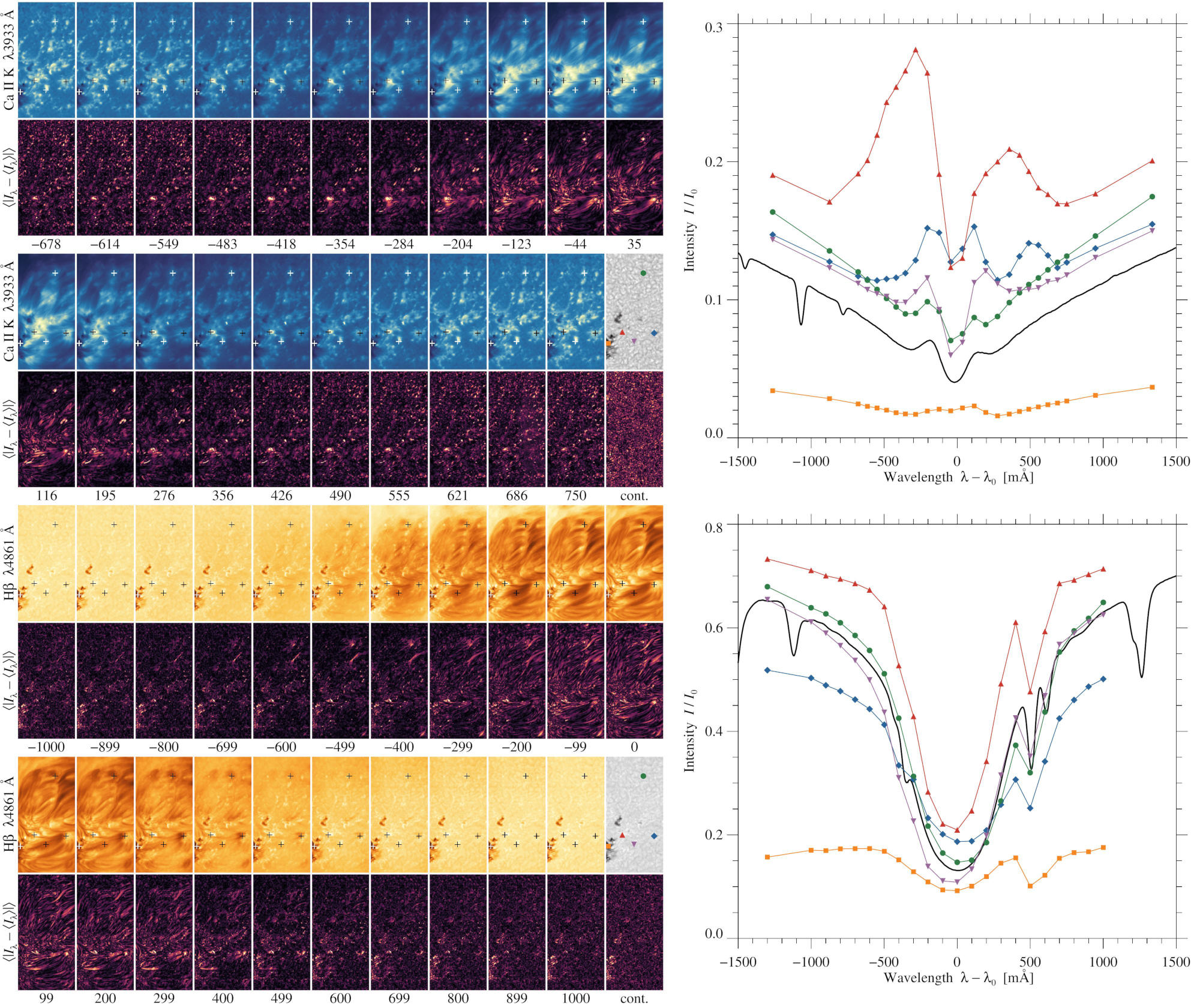}
\caption{Averaged spectral scans restored with MOMFBD of the 
    chromospheric \CaK (\textit{blue}) and \Hbeta (\textit{orange}) lines along with the corresponding spectral BaSAMs (\textit{plasma color table}). The wavelength positions $\lambda$ are given in milli{\aa}ngstr{\"o}ms as measured from the line center $\lambda_0$. The two grayscale maps represent the averaged restored continuum images. The two insets on the right show FTS atlas spectra (\textit{solid black}) with a spectral sampling of $\sim$2~m\AA\ of the \CaK and \Hbeta lines and CHROMIS spectra at five locations, where the positions are marked by color-coded symbols in the continuum images and by plus signs in the spectral scans.}
\label{FIG01}
\end{figure*}

\section{Observations}

High-resolution spectroscopic data were obtained with the CHROMospheric Imaging Spectrometer (CHROMIS) at the 1-meter Swedish Solar Telescope \citep[SST,][]{Scharmer2003} on La Palma, Canary Islands, Spain. The target was a small sunspot in the trailing part of active region NOAA~12723 from 08:18 to 09:14~UT on 2018 September~30, which was also observed by \citet{Kuckein2021} and \citet{Vissers2022}. The CHROMIS data are publicly available at the SST Data Archive\footnote{\url{https://dubshen.astro.su.se/sst_archive/observations/237}} after image restoration with multi-object multi-frame blind deconvolution \citep[MOMFBD,][]{vanNoort2005} and processing with the CRISPRED \citep{delaCruzRodriguez2015} and SSTRED \citep{Loefdahl2021} data reduction pipelines. The dataset consists of 205 spectral scans of the \CaK $\lambda$3933~\AA\ and \Hbeta $\lambda$4861~\AA\ lines with 26 and 23 wavelength points, respectively. The CHROMIS bandpass has a FWHM of about 12~pm, and the plate scale of the detectors is 0.38\arcsec~pixel$^{-1}$. Since image rotation is present, the common field-of-view (FOV) is significantly reduced to $28.9\arcsec \times 58.8\arcsec$ for the approximately 1-hour time series.

%===============================================================================
%   Methods
%===============================================================================

\section{Methods}

The background-subtracted variation of a spectral quantity $I_\lambda(t)$, that is, filtergrams or slit-reconstructed spectral maps, within a time series is given by
\begin{equation}
\begin{split}
\left\langle \left| I_\lambda - \left\langle I_\lambda \right\rangle \right| 
    \right\rangle & =  \frac{1}{N} \sum_{i=1}^N \left| I_\lambda(t_i) - 
    \left\langle I_\lambda \right\rangle \right|\\
\mathrm{with} \quad  \left\langle I_\lambda \right\rangle & = \frac{1}{N} 
    \sum_{i=1}^N I_\lambda(t_i),\\
\end{split}
\label{EQN01}
\end{equation}
where the subscript $\lambda$ refers to the wavelength position in a spectral line scan, and $t_i$ denotes the temporal position within the time series of spectral scans, and $N$ is the total number of scans \citep[cf.,][]{Denker2019}. Figure~\ref{FIG01} aggregates approximately $3.67 \times 10^{8}$ spectra in 26 \CaK and 23 \Hbeta background maps $\left\langle I_\lambda \right\rangle$ and BaSAMs $\left\langle \left| I_\lambda - \left\langle I_\lambda \right\rangle \right| \right\rangle$, where the angle brackets $\langle \ldots \rangle$ are shorthand for time averaging. Using 4$\times$4-pixel binning kept the plot window of Fig.~\ref{FIG01} to a manageable size.

%===============================================================================
%   Results
%===============================================================================

\section{Results}

Calculating spectral BaSAMs first produces an averaged spectral scan $\left\langle I_\lambda \right\rangle$ from a series of two-dimensional narrow-band filtergrams. In this illustrative example persistent spectral features were associated with sunspots, pores, and an arch filament system (AFS), connecting opposite magnetic polarities in the lower part of the FOV. This system is best seen in \Hbeta line-core filtergrams. Bright footpoints mark the ends of horizontal dark fibrils connecting the sunspot on the left to a small pore embedded in bright plages on the right. Such regions associated with small-scale magnetic flux elements are best seen in the far wings of the \CaK filtergrams, while \CaK line-core filtergrams indicate chromospheric heating near the sunspot and pores associated with emerging flux.

The \CaK line-wing BaSAMs $\left\langle \left| I_\lambda - \left\langle I_\lambda \right\rangle \right| \right\rangle$ clearly identify activity associated with the footpoints of the AFS, with the signal being stronger in the blue wing. In contrast, the right footpoint is more prominent in the \CaK line-core BaSAMs. The central filament shows strong absorption but no activity features in the BaSAMs. However, strong variations delineate this filament, suggesting filament oscillations or flows aligned with the filament axis.

In addition to the AFS, minor activity occurs near small-scale active-region filaments in the upper-right part of the FOV. These variations appear as a small kernel in the \CaK line-core BaSAMs and as filament-like features in the \Hbeta blue-wing and line-core BaSAMs. Overall, the morphology of the \CaK and \Hbeta BaSAMs falls into three categories related to the outer and inner line wings and the line cores, reflecting the transition from the photosphere to the chromosphere.

The potential of spectral BaSAMs is summarized in the two plots in Fig.~\ref{FIG01}, where \CaK and \Hbeta spectral lines for five locations are compared with the absolute disk-center intensity atlas spectrum \citep{Neckel1999} obtained with the McMath-Pierce Fourier Transform Spectrometer (FTS). Persistent intensity features in averaged spectral scans and regions with strong temporal variations, identified with spectral BaSAMs, served as objective criteria for selecting these locations. The averaged spectral profiles already show various spectral signatures, including blue- and red-shifts, line asymmetries, enhanced line wings, emission reversals, and (pseudo-)continuum and line-core intensities. These features, together with the averaged intensity and BaSAM values, can serve as a starting point for investigating the temporal evolution of spectra, for example, using machine learning techniques \citep[e.g.,][]{Verma2021}.

%===============================================================================
%   Acknowledgments
%===============================================================================

\noindent\begin{minipage}{\columnwidth}
\begin{acknowledgments}
\textit{\large Acknowledgments:} The CHROMIS data were obtained in a 2018 observing campaign by Joao da Silva Santos and Gregal Vissers. EU Horizon 2020 grant agreement 824135 (SOLARNET -- Integrating High-Resolution Solar Physics).
\end{acknowledgments}\vspace{-2mm}

\facilities{SST -- Swedish Solar Telescope, CHROMIS --  Chromospheric Imaging Spectrometer}

\software{BaSAMs \citep{Denker2019} -- MOMFBD \citep{vanNoort2005} -- 
   CRISPRED \citep{delaCruzRodriguez2015} -- SSTRED \citep{Loefdahl2021}}
\end{minipage}

%===============================================================================
%   Bibliography
%===============================================================================

\end{document}